\documentclass[10pt, twocolumn, article]{revtex4}

\usepackage{amsmath}
\usepackage{amssymb}

\input epsf
\input rotate

\def\bea{\begin{eqnarray}}
\def\eea{\end{eqnarray}}
\def\ben{\begin{equation}}
\def\een{\end{equation}}
\def\benu{\begin{enumerate}}
\def\enu{\end{enumerate}}

\def\lsim {\ifmmode {\buildrel<\over\sim}}

\def\t{\beta}

\def\1var{(\bx_1...\bx\N)}

\def\half{\frac{1}{2}}

\def\br{{\bf r}}

\def\b1{{\bf 1}}
\def\bx{{x}}

\def\N{_{\sss N}}

\def\sph_int{ {\int d^3 r}}

\def\infintd3r{ \int_{-\infty}^\infty d^3r\,}
\def\intd3r{ \int d^3r\,}

\def\laplace1d{\frac{d^2}{dx^2}}
\def\plaplace1d{\frac{d^2}{d{x'}^2}}

\def\padr2{\frac{\partial^2}{\partial r^2}}

\def\sV{{\cal V}}

\def\N{{\cal N}}
\def\a{{\alpha}}
\def\b{{\beta}}

\def\E{{\cal E}}
\def\G{{\cal G}}

\numberwithin{equation}{section}

\begin{document}


\title{{Partition theory: A very simple illustration}}
\author{Morrel H. Cohen}
\affiliation{Department of Physics and Astronomy, Rutgers
University, 126 Frelinghuysen Rd., Piscataway, NJ 08854, USA}
\affiliation{Department of Chemistry, Princeton University,
Washington Rd., Princeton, NJ 08544, USA}
\author{Adam Wasserman}
\affiliation{Department of Chemistry and Chemical Biology, Harvard
University, 12 Oxford St., Cambridge MA 02138, USA}
\author{Kieron Burke}
\affiliation{Department of Chemistry, University of California at
Irvine, 1102 Natural Sciences 2, Irvine, CA 92697, USA}

\begin{abstract}

We illustrate the main features of a recently proposed method
based on ensemble density functional theory to divide rigorously a
complex molecular system into its parts [M.H. Cohen and A.
Wasserman, J. Phys. Chem. A {\bf 111}, 2229 (2007)]. The
illustrative system is an analog of the hydrogen molecule for
which analytic expressions for the densities of the parts
(hydrogen ``atoms") are found along with the ``reactivity
potential" that enters the theory. While previous formulations of
Chemical Reactivity Theory lead to zero, or undefined, values for
the chemical hardness of the isolated parts, we demonstrate they
can acquire a finite and positive hardness within the present
formulation.

\end{abstract}

\maketitle

\section{Introduction}
In a series of recent papers \cite{CW03,CW06,CW07}, two of us have
developed a rigorous method for dividing a complex system into its
parts based on density-functional theory
\cite{HK64,KS65,L79,L82,PPLB82}. The underlying theory,
partition-theory (PT), was used to construct a formulation of
chemical reactivity theory (CRT) \cite{CW07} which, for the first
time, is consistent with the underlying density-functional theory
\cite{PPLB82,P85} and is richer in structure than the preexisting
CRT \cite{PY89,GPL03,PDLP78,PP83}.

In PT \cite{CW03,CW06,CW07}, a sharp definition of the individual
parts into which the whole system is partitioned is achieved first
by selecting the nuclei of each putative part and maintaining
these in the positions in which they occur in the whole and then
requiring that the sum of the electron densities of the parts,
each of which is treated as though isolated, add up exactly to the
electron density of the whole (the density constraint). The
electron densities of the parts are then to be determined by
minimizing the sum of the density functionals of the individual
parts with respect to the densities of the parts subject to the
density constraint. The density functional used, that of
ref.\cite{PPLB82} (PPLB), allows for the existence of noninteger
numbers of electrons on each part, necessary e.g. for the
definitions of electronegativity \cite{PDLP78} and hardness
\cite{PP83}, key indices of chemical reactivity \cite{CW07}, and
for incorporating covalent bonding between inequivalent parts.

The minimization proceeds via a Legendre transformation, which
introduces a reactivity potential $v_R(\br)$ as the Lagrange
multiplier of the density constraint. Thus, the formalism can
become computationally complex. First the electron density of the
whole system must be determined. Then, the densities of the parts
must be determined simultaneously with $v_R$, all of which is
required to set the stage for the determination of mutual
reactivities between parts, though certain self-reactivities can
be determined for each species alone without reference to a larger
system \cite{CW07}.

Accordingly, in the present paper, we develope the partition
theory in detail for an extremely simple system to exhibit its
main features explicitly. The illustrative system is an analog of
the hydrogen molecule in which the electrons move in one dimension
along the molecular axis without interacting, and the nuclear
Coulomb potentials are replaced by attractive delta-function
potentials. As a consequence of these extreme simplifications,
many quantities of interest can be determined analytically in a
transparent manner, including the electron density of the
molecule, of its parts (the ``atoms"), and the reactivity
potential at all internuclear separations.

In Section 2, the model is defined and the molecular density
obtained. In Section 3, the parts are defined, shown to have one
electron each, and a polar representation for their wave functions
found which facilitates the minimization. In Section 4, the
minimization is carried out, resulting in an Euler equation for
the polar angle $\beta(x)$ of that representation. $\beta(x)$ is
found in Section 5 and used to determine the reactivity potential
$v_R$ in Section 6. The principle of electronegativity
equalization formulated in refs.\cite{CW06} and \cite{CW07} is
shown to hold in Section 7. Also in Section 7, the hardness
\cite{CW07} of the isolated H atom is calculated, shown to be
nonzero, and correlated with the strength with which its electron
is bound. Thus, despite the fact that the model is a caricature of
the real system, meaningful features of the partition theory are
indeed illustrated by it, as discussed in the concluding Section,
8.

\section{1D-H$_2$; independent electrons moving in attractive $\delta$-function potentials in one dimension}
Our task is to partition an analog of the H$_2$ molecule in which
two electrons move independently  in $\delta$-function nuclear
potentials in one dimension into parts, analogs of H atoms. Each H
atom has, by symmetry, only one electron, so the need for the PPLB
density functional is avoided. Indeed no explicit use of
density-functional theory is required for either the molecule or
the atoms. The ground-state wave function $\psi_0$ and energy
$E_0$ of an isolated H atom are (atomic units are used
throughout):\begin{eqnarray}
\psi_0(x)&=&\sqrt{Z}e^{-Z|x|}~~,\label{e:psi_0}\\
E_0&=&-Z^2/2~~.\label{e:E_0}\end{eqnarray} In Eq.(\ref{e:psi_0}),
$(-Z)$ is the strength of the $\delta$-function potential. To draw
the analogy closer to real hydrogenic atoms, one could equate $Z$
to the nuclear charge.

The ground-state energy $E(N=1)$ of one electron moving
independently in the two $\delta$-function potentials centered at
$x=\pm a$ is $E(N=1)=-\kappa^2/2$, where $\kappa$ satisfies \ben
\kappa=2Z/(1+\tanh{\kappa a})\label{e:kappa}~~.\een The
corresponding wavefunction is:
\begin{eqnarray}\left.\begin{array}{ll}\psi_M(x)&=B e^{\kappa(a-|x|)}~~,~~|x|>a\\
&=B\frac{\cosh\kappa x}{\cosh \kappa
a}~~~~,~~|x|<a\end{array}\right\}~~,\label{e:psi_M}\end{eqnarray}
where \ben B=\kappa^{1/2}\left[1+\frac{\kappa a}{\cosh^2\kappa
a}+\tanh\kappa a\right]^{-1/2}~~;\label{e:B}\een Note that
$\kappa\to 2Z$ as $a\to 0$ (united atom limit) and $\kappa\to Z$
as $a\to\infty$ (separated atom limit).

The two-electron molecular electron density is given by: \ben
n_M(x)=2\left|\psi_M(x)\right|^2~~, \label{e:n_M}\een and the
total energy of the molecule is \ben
E_M(N=2)=2E_M(N=1)=-\kappa^2~~,\label{e:E_M}\een where $N$ is the
number of electrons in the molecule. The chemical potential of the
molecule is therefore \ben
\mu_M=E(2)-E(1)=E(1)=-\kappa^2/2~~.\label{e:mu_M}\een

\section{Parity decomposition}

We now partition the molecule into two parts $\a=1,2$, each having
a real one-electron wave function $\psi_\a$, localized around $-a$
and $+a$ respectively, so that $n_M(x)$ is given by \ben
n_M(x)=n_1(x)+n_2(x)~~,\label{e:density_constraint}\een where
$n_\a(x)$ is the electron density of each part $\a=1,2$ treated
independently. The ``atomic" wavefunctions are given by: \ben
\psi_\a(x)=\sqrt{n_\a(x)}~~.\label{e:psi_a}\een They are mirror
images of each other, \ben
\psi_2(x)=\psi_1(-x)~~,\label{e:mirror}\een and both are
normalized.

We now decompose the $\psi_\a$ into their symmetric,
$\psi_s(-x)=\psi_s(x)$, and antisymmetric,
$\psi_a(-x)=-\psi_a(x)$, parts by a rotation within the function
space they span, \begin{eqnarray}
\psi_1=\frac{1}{\sqrt{2}}\left(\psi_s+\psi_a\right)~~,
~~\psi_2=\frac{1}{\sqrt{2}}\left(\psi_s-\psi_a\right)~~;\label{e:psi1psi2}\\
\psi_s=\frac{1}{\sqrt{2}}\left(\psi_1+\psi_2\right)~~,
~~\psi_a=\frac{1}{\sqrt{2}}\left(\psi_1-\psi_2\right)~~.\label{e:psispsia}
\end{eqnarray}
The rotation leaves ``lengths" within the space invariant so that
\ben n_M=\psi_s^2+\psi_a^2~~.\een We next introduce
$\beta=\beta(x)$, a polar angle in the function space, \ben
\psi_s=\sqrt{n_M}\cos\t~~,~~\psi_a=\sqrt{n_M}\sin\t~~,\label{e:parity_for_psispsia}\een
so that \ben \psi_{1,2}=\sqrt{n_M/2}\left(\cos\t\pm\sin\t\right)
\label{e:parity}\een Because the $\psi_\a$ are non-negative,
$|\t|$ cannot exceed $\pi/4$. Furthermore $\beta$ must be an odd
function of $x$, to ensure $\psi_\a$ is also odd. This also
guarantees normalization of $\psi_\a$.

\section{The Euler equation for $\t(x)$}

To apply PT \cite{CW06,CW07}, begin with the original Hamiltonian
\ben
H=\half\sum_{i=1,2}p_i^2-Z\sum_{i=1,2}\left[\delta(x_i-a)+\delta(x_i+a)\right]~~.\label{e:original_H}\een
Then divide the system into overlapping regions, each with a given
number of electrons. In this case, we choose one electron on the
left, and the other on the right. Thus we have two 1-electron
problems: \ben H_\a=\frac{p^2}{2}+v_\a~~,~~ v_{1,2}=-Z\delta(x\mp
a)~~.\label{e:Hparts}\een The PT problem is to minimize
\ben\E=\left(\psi_1,H_1\psi_1\right)+\left(\psi_2,H_2\psi_2\right)~~,\label{e:functional}\een
subject to normalization of the wavefunctions, but also to the
constraint that the total density equal the original molecular
density, Eq.(\ref{e:density_constraint}). (Without the latter
constraint, we'd obviously find $\psi_{1,2}=\psi_0(x=\mp a)$).
 In the polar representation of Sec.3, both density and
normalization constraints are automatically satisfied, so the
partition problem becomes simply minimizing $\E$ as a functional
of $\t$. That functional is
\begin{eqnarray}\nonumber \E=&&\int{
dx\left\{\frac{1}{2}\left[\frac{1}{4}\frac{n_M'^2}{n_M}-\half
n_M''+n_M(\t')^2\right]\right.}\\&&~~+\left.\half
n_M\left[(v_1+v_2)+(v_1-v_2)\sin
2\t\right]\right\}.\end{eqnarray} Varying it yields \ben
\delta\E=\int dx{\left\{n_M\t'\delta \t'+(v_1-v_2)n_M\cos
2\t\delta\t\right\}}~~.\een Integrating by parts, as usual, leads
to \begin{eqnarray}\nonumber
\left.\delta\E=2n\t'\delta\t\right|_{x=-\infty}^{x=+\infty}+\int
dx{\left\{\frac{d}{dx}\left(n_M\frac{d\t}{dx}\right)\right.}\\\left.
~~~~+(v_1-v_2)n_M\cos
2\t\right\}\delta\t~~.\label{e:int_parts}\end{eqnarray} For $\E$ to
be stationary with respect to arbitrary variations $\delta\t$ of
$\t$, both terms contributing to $\delta\E$ in
Eq.(\ref{e:int_parts}) must vanish. The Euler equation which
results from the vanishing of the second term in
Eq.(\ref{e:int_parts}) is
\begin{eqnarray}
-\frac{d}{dx}\left(n_M\frac{d\t}{dx}\right)+(v_1-v_2)n_M\cos
2\t=0~~,\\\nonumber
\frac{d}{dx}\left(n_M\frac{d\t}{dx}\right)+
Z\left(\delta(x-a)-\delta(x+a)\right)\times\\\times n_M\cos
2\t=0~~.\label{e:Euler}\end{eqnarray} The vanishing of the first
term in Eq.(\ref{e:int_parts}) sets the boundary condition at
infinity on the Euler equation (\ref{e:Euler}). There are two
possibilities, the vanishing of $\t'$ at infinity or the fixing of
$\t$ there so that $\delta\t$ must vanish. As we shall see in
Section 5, imposing the latter results in an unacceptable
divergence in $\t'$ at infinity. We therefore impose the boundary
condition \ben \t'(x)=0~~,~~|x|=\infty~~.\label{e:boundary}\een

\section{Solving for $\t(x)$}
Eq.(\ref{e:Euler}) becomes \ben
\frac{d}{dx}\left(n_M\frac{d\t}{dx}\right)=0~~,~~|x|\neq
a~~,\label{e:beta_eq}\een subject to the boundary conditions
Eq.(\ref{e:boundary}) and
\begin{eqnarray}
\left.\begin{array}{ll}\t(a^-)=\t(a^+)\equiv\t_a\\\t'(a^-)-\t'(a^+)=Z\cos
2\t_a\end{array}\right\}&x&=a~,\label{e:r-bound}\\
\left.\begin{array}{ll}\t(-a^+)=\t(-a^-)=-\t_a\\\t'(-a^-)-\t'(-a^+)=Z\cos
2\t_a\end{array}\right\}&x&=-a~.\label{e:l-bound}
\end{eqnarray}
The general solution of (\ref{e:beta_eq}) is \begin{eqnarray}
\frac{d\t(x)}{dx}&=&\frac{c_1}{n_M(x)}~~,\label{e:general}\\\t(x)&=&\int^x
dx'\frac{c_1}{n_M(x)}+c_2~~.\label{e:beta_general}\end{eqnarray}
where $c_1$ and $c_2$ are constants. As implied above in Section
4, if $c_1$ does not vanish $\t'$ diverges exponentially at
infinity, according to Eq.(\ref{e:general}), because $n_M$ goes
exponentially to zero, so, in accordance with
Eq.(\ref{e:boundary}), $c_1$ vanishes for $|x|>a$, and $\t(x)$ is
constant there, \ben\left.
\begin{array}{ll}\t(x)&=\t_a~~~~~~~,~~x>a\\&=-\t_a~~~~,~~x<-a\end{array}\right\}~~.\label{e:beta_asympt}\een
For $|x|<a$ we can rewrite Eq.(\ref{e:beta_asympt}) as \ben
\t(x)=\int_{-a}^x
dx'\frac{c_1}{n_M(x')}-\t_a~~,\label{e:beta_int}\een which implies
that \ben \t_a=\half\int_{-a}^a dx
\frac{c_1}{n_M(x)}~~.\label{e:beta_a}\een From (\ref{e:beta_a}) we
can relate $c_1$ to $\t_a$ via Eq.(\ref{e:n_M}), \ben
c_1=\frac{2\kappa B^2\t_a}{\cosh^2{\kappa a}\tanh{\kappa
a}}~~.\label{e:c1}\een Inserting (\ref{e:c1}) for $c_1$ into
Eq.(\ref{e:general}) and the result into the BC (\ref{e:r-bound})
or (\ref{e:l-bound}) produces an equation for $\t_a$, \ben
\t_a=\frac{Z}{2\kappa}\sinh{2\kappa
a}\cos{2\t_a}~~.\label{e:betaofa}\een Inserting Eqs.(\ref{e:c1})
and (\ref{e:n_M}) into Eq.(\ref{e:beta_int}) yields the remarkably
simple result \ben \t(x)=\frac{\tanh{\kappa x}}{\tanh{\kappa a}}
\beta_a~~,~~0<|x|<a~~.\label{e:betax}\een

\begin{figure}
\begin{center}
\epsfxsize=70mm \epsfbox{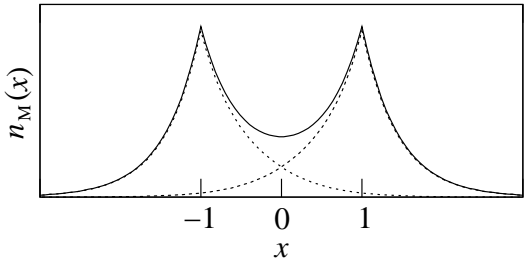}
\caption{Molecular density $n_M(x)$ (solid), and ``atomic"
densities $n_1(x)$ and $n_2(x)$ (dotted) for $Z=1$ and $a=1$.
}\label{f:fig1a}
\end{center}
\end{figure}

\begin{figure}
\begin{center}
\epsfxsize=70mm \epsfbox{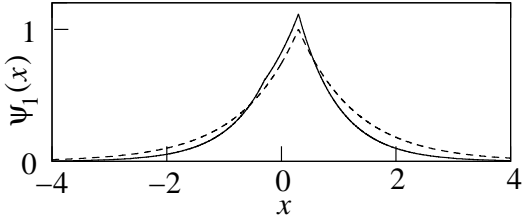}
\caption{Right-side ``atomic" orbital $\psi_1(x)$ (solid) and pure
exponential orbital $\psi_0(x)$ (dashed) for $Z=1$ and
$a=0.3$.}\label{f:fig1b}
\end{center}
\end{figure}

\begin{figure}
\begin{center}
\epsfxsize=70mm \epsfbox{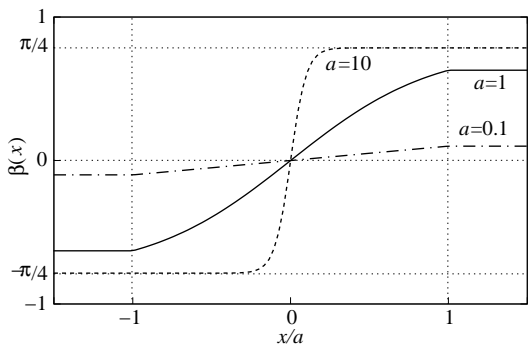}
\caption{$\t(x)$ {\em vs.} $x$, as given by Eq.(\ref{e:betax}),
for fixed $Z=1$ and 3 different values of $a$.}\label{f:fig2}
\end{center}
\end{figure}

Eqs.(\ref{e:beta_asympt}), (\ref{e:betaofa}), and (\ref{e:betax}),
together with Eq.(\ref{e:kappa}) provide a complete analytic
solution for $\t(x)$ and through Eqs.(\ref{e:psispsia}) and
(\ref{e:parity}) for the $\psi_\a$. In Fig.{\ref{f:fig1a}} we show
$n_M$, $n_1$ and $n_2$ {\em vs.} $x$ for $Z=1$ and $a=1$. We see
that each localized density spreads into the neighboring region, and
looks quite similar to an atomic density. To see the differences
from isolated atomic orbitals, in Fig.{\ref{f:fig1b}} we make the
distance smaller ($a=0.3$), and show the right-side ``atomic"
orbital $\psi_1(x)$ (solid line) and compare it with the pure
exponential orbital $\psi_0(x)$ of Eq.{\ref{e:psi_0}} (dashed
line). The orbital $\psi_1$ resembles $\psi_0$ and tends to it for
large $a$, but is distorted with respect to it for small $a$. Its
maximum is still a cusp at $x=a$, but it also shows a second cusp
at $x=-a$. Since $\kappa>Z$ always (Eq.(\ref{e:kappa})), and
either $\psi_1$ or $\psi_2$ is proportional to $\psi_M$ for
$|x|>a$, where $\t=\t_a$ is constant, the PT atomic densities and
orbitals decay more rapidly than isolated atoms. Since their
normalization is the same, this in turn means enhanced density
between the `nuclei', due to bonding. In Fig.{\ref{f:fig2}}, we
show $\t(x)$ for $Z=1$, and $a=0.1$, $1$, and $10$. Qualitatively,
from Eq.(\ref{e:betax}),
\begin{eqnarray}\nonumber \t(x)&\simeq& \t_a\frac{x}{a}~~,~~x<{\rm min}(1/\kappa,a)\\&=&\t_a~~~~,~~
x>{\rm min}(1/\kappa,a)\nonumber\end{eqnarray} and if $Za>>1$
(large separation), $\t_a\simeq\pi/4$ while if $Za<<1$ (small
separation), $\t_a\simeq a$. The interpretation of these results
is given in terms of (\ref{e:parity}), outside the bond region. If
$\t_a$ is small, both `atoms' share the density in each outside
region. But if $\t_a$ is close to $\pi/4$, each atom dominates on
its own side, consuming the entire density there.


\section{The reactivity potential}

\begin{figure}
\begin{center}
\epsfxsize=40mm \epsfbox{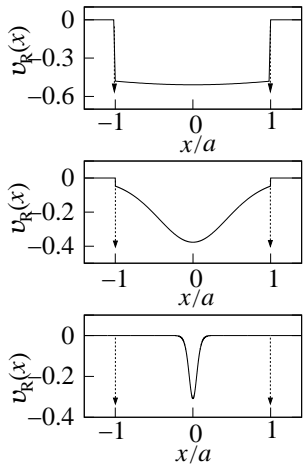}
\caption{Reactivity potential $v_R$, Eq.(\ref{e:6.9}) for fixed
$Z=1$ and 3 different values of $a$: $a=0.1$ (upper panel), $a=1$
(middle) and $a=10$ (bottom). The $\delta$-functions at $\pm a$
are indicated by arrows.} \label{f:vR}
\end{center}
\end{figure}

\begin{figure}
\begin{center}
\epsfxsize=70mm \epsfbox{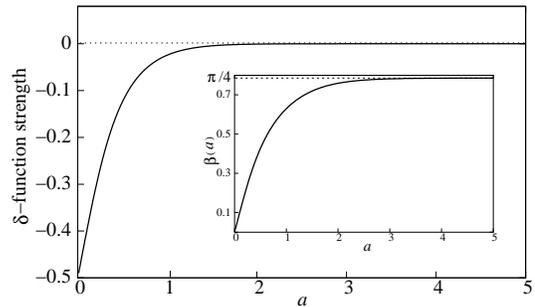}
\caption{Weights of the $\delta$-function components of $v_R$
divided by $Z$ as a function of $a$ for fixed $Z=1$, from the
second term of Eq.(\ref{e:6.9}). The inset shows $\t_a$ vs. $a$}
\label{f:weights}
\end{center}
\end{figure}

The one-electron wave functions $\psi_1(x)$ and $\psi_2(x)$ are
not eigenstates of the part-Hamiltonians $H_1$ and $H_2$ of
Eq.(\ref{e:Hparts}). The natural question arises: What are they
eigenstates of? The partition theory of refs.
\cite{CW03,CW06,CW07} dictates that they are eigenstates of the
modified single-electron Hamiltonians $H_\a^{R}=p^2/2+\sV_\a$,
$\a=1,2$:
\begin{eqnarray}
\left(\frac{p^2}{2}+\sV_\a\right)\psi_\a&=&\mu_M\psi_\a~~,~~\a=1,2~~.\label{e:6.1}\\
\sV_\a&=&v_\a+v_R\label{e:6.1a}
\end{eqnarray}
where the eigenvalue, regardless of the part $\a$, is precisely
equal to the molecular chemical potential $\mu_M$ of
Eq.(\ref{e:mu_M}). The potential $v_R(x)$ is the {\em reactivity
potential} that we now construct explicitly. Summing over $\a$ and
dividing by $\psi_1+\psi_2$ yields a symmetric expression for
$v_R$, \ben
v_R=\mu_M-\frac{1}{\psi_1+\psi_2}\frac{p^2}{2}\left(\psi_1+\psi_2\right)
-\frac{v_1\psi_1+v_2\psi_2}{\psi_1+\psi_2}~~.\label{e:6.2}\een
$\psi_1$ and $\psi_2$ can be reexpressed in terms of $\psi_s$ and
$\psi_a$, Eq.(\ref{e:psispsia}). Noting that \ben
n_M=2\psi_M^2~~,\label{e:6.3}\een using
Eq.(\ref{e:parity_for_psispsia}) for $\psi_{s,a}$, and taking the
$\delta$-function character of $v_\a$ into account results in
\begin{eqnarray}\nonumber
v_R&=&\mu_M+\frac{1}{2\psi_M\cos\t}\frac{d^2}{dx^2}\left(\psi_M\cos\t\right)\\
&~&-\half(v_1+v_2)(1+\tan\t_a)~~.\label{e:6.4}\end{eqnarray} The
molecular wave function $\psi_M$ satisfies the Schr\"{o}dinger
equation, \ben
-\frac{1}{2}\frac{d^2\psi_M}{dx^2}+(v_1+v_2)\psi_M=\mu_M\psi_M~~,\label{e:6.5}\een
which can be used to transform Eq.(\ref{e:6.4}) to
\begin{eqnarray}\nonumber v_R&=&-\frac{1}{2}\left\{\left[\frac{2}{\psi_M}\frac{d\psi_M}{dx}
\frac{d\t}{dx}+\frac{d^2\t}{dx^2}\right]\tan\t+\left(\frac{d\t}{dx}\right)^2\right\}\\
&~&+\half(v_1+v_2)(1-\tan\t_a)~~.\label{e:6.6}\end{eqnarray} Using
Eq.(\ref{e:6.3}), the Schr\"{o}dinger-like equation for $\t$,
Eq.(\ref{e:Euler}), can be rewritten as
\begin{eqnarray}\nonumber
&-&\frac{1}{2}\left[\frac{2}{\psi_M}\frac{d\psi_M}{dx}\frac{d\t}{dx}+\frac{d^2\t}{dx^2}\right]\\
&~~&+\half(v_1-v_2)\cos 2\t_a=0~~.\label{e:6.7}\end{eqnarray}
Multiplying Eq.(\ref{e:6.7}) by $\tan\t$, invoking the oddness of
$\t$ and the $\delta$-functions in $v_1$ and $v_2$, and
subtracting the result from Eq.(\ref{e:6.6}) yields for $v_R$ \ben
v_R=-\frac{1}{2}\left(\frac{d\t}{dx}\right)^2+\half(v_1+v_2)\left[1-(1+\cos
2\t_a)\tan\t_a\right]~~.\label{e:6.8}\een Inserting our previous
result for $\t(x)$, Eqs.(\ref{e:beta_asympt}) and (\ref{e:betax})
into (\ref{e:6.8}) yields an explicit result for $v_R$,
\begin{eqnarray}\nonumber v_R&=&\frac{\mu_M\t_a^2}{\tanh^2{\kappa
a}}\frac{\theta(a-|x|)}{\cosh^4{\kappa x}}\\&~~&+
\half(v_1+v_2)\left[1-\sin{2\t_a}\right]~~,\label{e:6.9}\end{eqnarray}
where $\theta(y)=0$ for $y<0$, $1$ for $y>0$ is the Heaviside step
function. Eq.(\ref{e:6.9}) shows that $v_R(x)$ vanishes for
$|x|>a$, has attractive $\delta$-functions at $\pm a$ whose
weights increase monotonically from $0$ to $\half Z$ as $Za$
decreases from infinity to zero, and has an attractive inverse
cosh$^4(x)$ component for $|x|<a$. For the united atom case,
$Za\downarrow 0$, $v_1+v_R=v_2+v_R=2v_1$ simply reproduces the
molecular potential, and $\psi_1=\psi_2=\psi_M$ as they should.
Figure \ref{f:vR} displays $v_R$ {\em vs.} $x$ for fixed $Z=1$ and
representative values of $a$. The reactivity potential is almost
flat for small separations, a wide well in between the two atoms
for intermediate separations, and a narrow well that is far from
both atoms at large separations. Figure \ref{f:weights} displays
the weights of the $\delta$-function components of $v_R$ divided
by $Z$ vs. $a$.

As shown in ref.\cite{CW07}, the Kohn-Sham (KS) HOMO eigenvalue of
each part must be identical to the chemical potential of the whole
in the added presence of $v_R$. In our simple example, the KS
potential of a part reduces to the nuclear $\delta$-function
potential of one H atom. Adding $v_R$ to the nuclear potential
must therefore transform the HOMO energy $E_0$, Eq.(\ref{e:E_0}),
of the isolated atom to the more negative HOMO energy of the
molecule $E(N=1)=-\kappa^2/2$, which is its chemical potential
(Eq.(\ref{e:mu_M})). $v_R$ must be attractive to do that, which it
is, from Eqs.(\ref{e:6.8}) and (\ref{e:6.9}). In our simple
example, $v_R$ makes the delta function of the atom more negative,
adds the attractive inverse cosh$^4$ potential between the atoms,
and adds an attractive ghost delta function at the position of the
other atom to force the wave function to decay sufficiently
rapidly outside the molecule.

\begin{figure}
\begin{center}
\epsfxsize=60mm \epsfbox{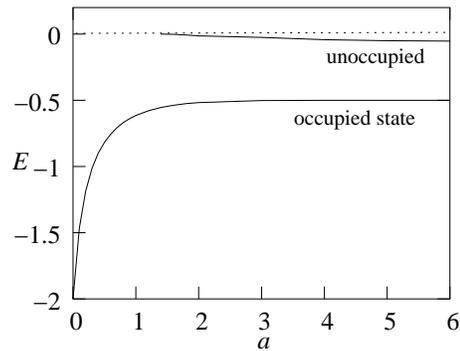}
\caption{Energy as a function of $a$, in atomic units, for the two
lowest-energy solutions of Eq.(\ref{e:6.1}). $Z=1$ for this
plot.}\label{f:ghost}
\end{center}
\end{figure}

In the limit of infinite separation $v_1+v_R$ reduces to $v_1$ and
$v_2+v_R$ reduces to $v_2$, except for $|x|<a$, where the
attractive potential \ben
v_R(x)=\frac{\pi^2E_0}{16}\frac{1}{\cosh^4{Zx}}~~,~~|x|<a~~,\label{e:6.11}\een
persists. This potential has at least one additional bound state,
but with binding energy less than $|E_0|$. Thus it is unoccupied,
and does not affect our results.
The $a$-dependence of this state's energy is shown for fixed $Z$
in Fig.\ref{f:ghost}.
For very large separation between the atoms, it is localized at
the center of the inverse cosh$^4(x)$ component of $v_R$, but it
rapidly delocalizes for smaller separations. In particular, for
$Z=1$, it is highly delocalized when $a<\sim 1.4$, where it
vanishes into the continuum.

\section{Susceptibility and hardness}

Having found the reactivity potential, we now illustrate the
construction of reactivity indices. In the CRT of ref.\cite{CW07},
each part $\a$ is represented by an ensemble of PPLB type
containing contributions with only two integer electron numbers,
$p_\a$ and $p_{\a+1}$. The principle of electronegativity
equalization is expressed as the equality of the chemical
potential of each part in the presence of the reactivity
potential, $\mu_\a^R$, to the chemical potential of the molecule,
$\mu_M$, \ben \mu_\a^R=\mu_M~~,~~\forall\a~~.\label{e:a}\een The
$\mu_\a^R$ are defined as the difference between the ground state
energies of $\a$ for $p_\a+1$ and $p_\a$ electrons in the presence
of $v_R$, \ben \mu_a^R=E_\a^R(p_\a+1)-E_\a^R(p_\a)~~,\een and
similarly for $\mu_M$ \ben \mu_M=E_M(N_M)-E_M(N_M-1)~~.\een

In our simple example, $\mu_M$ is given in Eq.(\ref{e:mu_M}). The
relevant value of $p_\a$ is zero, so that $\mu_\a^R$ is just
$E_\a^R(1)$, the lowest eigenvalue of \ben
H_\a^R=H_\a+v_R~~,\label{e:d}\een with $H_\a$ given by
Eq.(\ref{e:Hparts}) and $v_R$ by Eq.(\ref{e:6.9}). The explicit
construction of $v_R$ in Section 6, not possible in general,
guarantees that Eq.(\ref{e:a}) and therefore electronegativity
equalization holds. In the general case, a modification of the
Car-Parrinello scheme \cite{CP85,CCunp} guarantees
electronegativity equalization.

\begin{figure}
\begin{center}
\epsfxsize=80mm \epsfbox{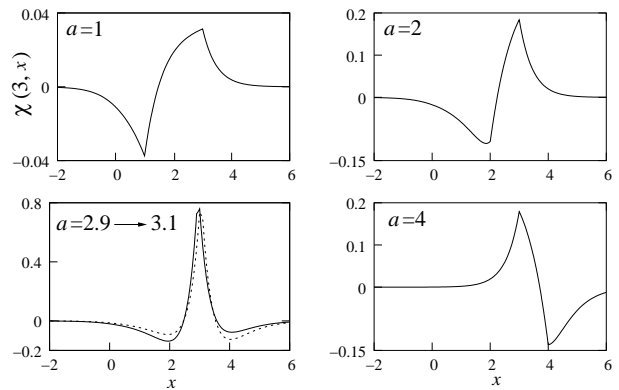}
\caption{Susceptibility $\chi(x_0,x)$ of the right ``atom"
obtained from Eqs.(\ref{e:7.2})-(\ref{e:7.4}), as indicated in
the Appendix, when $x_0$ is set to 3 a.u. Each panel corresponds to
a different value of the internuclear distance, $a$. The
lower-left panel shows $\chi(x_0,x)$ when $a$ is just below
(solid) and just above (dotted) $x_0$.
 } \label{f:suscep}
\end{center}
\end{figure}

The susceptibility of part $\a$ measures the response of the
density of part $\a$ to a small change in the potential ${\cal
V}_\a$ of Eq.(\ref{e:6.1a}): \ben \chi_\a(x,x')=-\frac{\delta
n_\a(x)}{\delta {\cal V_\a}(x')}~~.\label{e:7.1}\een For 2
electrons, it is simple to show that \ben
\chi_\a(x,x')=-2\psi_\a(x)\G_\a(\mu_M;x,x')\psi_\a(x')~~,\label{e:7.2}\een
where $\G_\a(\mu_M;x,x')$ is given by the $E\to\mu_M$ limit of:
\ben
\G_\a(E;x,x')=G_\a(E;x,x')-\frac{\psi_a(x)\psi(x')}{E-\mu_M}~~,\label{e:7.3}\een
and $G_\a$ is the Green's function for part $\a$: \ben
G_\a(E;x,x')=\left[E-\left(\frac{p^2}{2}+\sV_\a\right)\right]^{-1}(x,x')~~.\label{e:7.4}
\een Figure \ref{f:suscep} shows the susceptibility of the right
``atom" for various interatomic separations when the perturbing
potential is added at $x_0=3$ (the numerical calculations were
done as described in the Appendix). Electrons flow away from $x_0$,
building up a peak at $x_0$ (positive because of the minus sign in
the definition of $\chi_\a$, Eq.(\ref{e:7.1})), and a negative
peak at the closest maximum of the charge density, i.e. at $a$.
With the analytic Green function of an isolated ``atom"\cite{SO89}
and Eqs.(\ref{e:7.2})-(\ref{e:7.3}), $\chi_\a$ can be obtained
analytically in the large-separation limit:
\begin{eqnarray}\nonumber
\chi_\a(x,x')=2e^{-Z|x|}\left\{e^{-Z|x-x'|}-\left[\half+Z\left(|x|+|x'|\right)\right]\right.\\
\times\left.e^{-Z(|x|+|x'|)}\right\}e^{-Z|x'|}~~~~~ \end{eqnarray}

\begin{figure}
\begin{center}
\epsfxsize=60mm \epsfbox{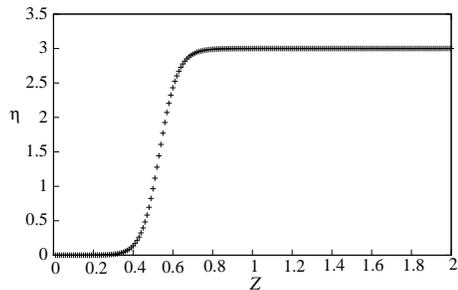}
\caption{Self-hardness vs. $Z$ in the separated-atom limit (atomic
units).} \label{f:hardness}
\end{center}
\end{figure}

We now construct the susceptibility of the whole system, $\chi_R$,
by adding together the susceptibilities of the parts, \ben
\chi_R(x,x')=\sum_\a\chi_\a(x,x')~~.\label{e:7.5}\een The inverse
of $\chi_R$ determines the hardness matrix $\eta_{\a\b}$ as shown
in refs.\cite{CW06} and \cite{CW07}: \ben \eta_{\a\b}=\int\int dx
dx'f_\a(x)\chi_R^{-1}(x,x')f_\b(x')~~,\label{e:7.6}\een where the
Fukui function of part $\a$, $f_\a(x)$, \ben
f_\a(x)=\frac{dn_\a(N_a,x)}{dN_a}~~, \label{e:7.7}\een is simply
equal to $\psi_\a^2(x)$ for 2 non-interacting electrons, since
$n_\a(N_\a,x)=N_\a\psi_\a^2(x)$ (see also ref.\cite{FCC07}). Thus,
we have \ben \eta_{\a\b}=\int\int dx
dx'\psi_\a^2(x)\chi_R^{-1}(x,x')\psi_\b^2(x')~~.\label{e:7.8}\een
Figure \ref{f:hardness} shows the self-hardness $\eta_{\a\a}$ for
an isolated H-``atom", as a function of $Z$. The constancy of the
hardness for large $Z$ can be understood qualitatively as follows.
The inverse susceptibility has units of energy times length
squared. When $Z$ is large, it establishes a length scale
inversely proportional to $Z$, and an energy scale proportional to
$Z^2$, so the $Z$-dependence cancels out in the inverse
susceptibility. To obtain the hardness, we multiply $\chi_R^{-1}$
on the left and right by the Fukui function, which has the
dimension of inverse length. Integrating over position on the left
and right then cancels out the $Z$-dependence arising from the
Fukui functions, and the result is a $Z$-independent hardness.

\section{Conclusions}

Despite the extreme simplicity of the 1D-H$2$ model analyzed here
-- two non-interacting electrons moving in 1D under the influcence
of two equivalent attractive delta-function potentials -- that
model allows us to illustrate the essential features of our
partition theory and of key indices of our chemical reactivity via
straightforward analysis and easy computations.

We have shown that the electron density of the molecule can be
decomposed exactly into a sum of atomic densities, a rigorous
solution of the ``atoms-in-molecules" problem \cite{AIM}.

Electronegativity equalization \cite{S51} is built into the
partition by the symmetry of the problem, so this homonuclear
model does not illustrate that principle as well as a
heteronuclear model would. Nevertheless, the current example does
illustrate a key feature of the new CRT, the chemical context
dependence of the reactivity indices, in this case the
electronegativity of a part, introduced through the presence of
$v_R$ in the Schr\"{o}dinger equation for $\psi_\a$, cf.
Eq.(\ref{e:d}). It also demonstrates that the reactivity potential
remains finite as two atoms separate, but has no effect on the
partitioning after separation.

Another serious shortcoming of the earlier formulations of
DFT-based CRT is the vanishing of the hardness. We have shown
explicitly here that the self-hardness, as defined in \cite{CW07},
of an isolated ``atom" is positive. Interestingly, the hardness
saturates as the ionization energy of the ``atom" increases,
raising the very interesting question of whether such a
saturation of hardness with ionization energy exists in real
systems.
For this model, a strong positive correlation between hardness and ionization energy exists only over the limited range of $Z$ between 0.4 and 0.7.

\vspace{0.5cm} KB is supported by NSF CHE-0355405.

\renewcommand{\theequation}{A.\arabic{equation}}
\setcounter{equation}{0}  
\section*{APPENDIX: Numerical calculation of the susceptibility}  
We first obtained $G_\a(E;x,x')$ according to the well-known
prescription \cite{A70}: \ben
G_\a(E;x,x')=2\frac{\psi_{\a,L}(E,x_<)\psi_{\a,R}(E,x_>)}{W[\psi_{\a,L},\psi_{\a,R}]}~~,
\een where $x_<={\rm inf}(x,x')$, $x_>={\rm sup(x,x')}$
\begin{eqnarray}\nonumber W[\psi_{\a,L},\psi_{\a,R}]=
&&\psi_{\a,L}(E,x)\psi_{\a,R}'(E,x)\\&&-\psi_{\a,L}'(E,x)\psi_{\a,R}(E,x)~,\end{eqnarray}
and the orbitals $\psi_{\a,L}$ and $\psi_{\a,R}$ are solutions of
\ben
\left[\frac{p^2}{2}+\sV_\a(x)\right]\psi_{\a,L,R}(E,x)=E\psi_{\a,L,R}(E,x)\label{e:A.3}\een
satisfying left and right-boundary conditions, respectively:
\begin{eqnarray} |\psi_{\a,L}(E,x)|\downarrow
0~~&,&~~x\downarrow-\infty\\|\psi_{\a,R}(E,x)|\downarrow
0~~&,&~~x\uparrow\infty~~\end{eqnarray} The potential $\sV_\a(x)$
of Eq.(\ref{e:A.3}) is given by Eq.(\ref{e:6.1a}), with the
reactivity potential $v_R(x)$ of Eq.(\ref{e:6.9}). The
computations of $\psi_{\a,L,R}(E,x)$ were carried out at
$E=\mu_M\pm\Delta E$ with $\Delta E$ chosen for numerical
convenience, i.e. large enough so that ${\rm
sup}_{x,x'}|G_\a(\mu_M\pm\Delta E)|$ does not become so large as
to be inconvenient on the one hand, and small enough so that
$\half\left[G_\a(\mu_M+\Delta E)+G_\a(\mu_M-\Delta E)\right]$ does
not differ significantly from its limit at $\Delta E\downarrow 0$.
We then calculated $\G_\a$ of Eq.(\ref{e:7.3}) as: \ben
\G_\a(\mu_M;x,x')=\half\left[G_\a(\mu_M+\Delta E;
x,x')+G_\a(\mu_M-\Delta E; x,x')\right]~~\een \vspace{3cm}

\end{document}